# Fault Localization for Declarative Models in Alloy

Kaiyuan Wang, Allison Sullivan[†], Darko Marinov[*], and Sarfraz Khurshid

The University of Texas at Austin, [†]North Carolina A&T State University, [*]University of Illinois at Urbana-Champaign
{kaiyuanw,khurshid}@utexas.edu,allisonksullivan@gmail.com,marinov@illinois.edu

## ABSTRACT

Fault localization is a popular research topic and many techniques have been proposed to locate faults in imperative code, e.g. C and Java. In this paper, we focus on the problem of fault localization for declarative models in Alloy – a first order relational logic with transitive closure. We introduce AlloyFL, the first set of fault localization techniques for faulty Alloy models which leverages multiple test formulas. AlloyFL is also the first set of fault localization techniques at the AST node granularity. We implements in AlloyFL both spectrum-based and mutation-based fault localization techniques, as well as techniques that are based on Alloy's built-in unsat core. We introduce new metrics to measure the accuracy of AlloyFL and systematically evaluate AlloyFL on 38 real faulty models and 9000 mutant models. The results show that the mutation-based fault localization techniques are significantly more accurate than other types of techniques.

## 1 INTRODUCTION

Writing declarative models and specifications has numerous benefits, ranging from automated reasoning and correction of design-level properties before systems are built [16, 26], to automated testing and debugging of their implementations after they are built [36]. However, correctly writing declarative models that represent nontrivial properties is not easy, especially for practitioners who are not well-versed with the intricate syntax and semantics of declarative languages. Our focus in this paper is declarative models written in Alloy [16] – a first-order relational logic with transitive closure. We choose Alloy because of its expressive power and use in numerous domains like security [33, 41], networking [50], UML analysis [34, 35], etc. The Alloy Analyzer provides an automatic analysis engine for Alloy based on off-the-shelf SAT solvers [10] and it is able to generate valuations for the relations in the models such that the properties modeled hold or are refuted as desired.

Existing Alloy users typically write formulas and commands to check if the model complies for a set of expected properties. For example, Pamela Zave invokes a set of Alloy predicates and assertions in her model [1] to check the expected properties of the Chord [54] distributed hash table protocol. We refer to these Alloy predicates, functions and assertions that check the expected model properties as Alloy "*tests*" in the rest of this paper. These tests can help capture modeling errors and regression errors analogous to tests in imperative languages like Java. Existing debugging techniques in Alloy, e.g. MiniSat solver with unsat core [53], highlight suspicious code snippets for a *single* test.

To improve the debugging process, we introduce AlloyFL, the first set of fault localization (FL) techniques that leverage *multiple* tests for declarative models written in Alloy. While our focus is Alloy, our techniques can be generalized to other languages that do not have imperative notions of control flow or statements. Our key insight is that a *test-driven* approach inspired by traditional FL based on passing and failing tests for *imperative* code [4, 6, 15, 17, 19, 20, 24, 39, 40, 44, 46, 51, 62, 66, 67] can also lay the foundation for effective localization of faults in declarative models. Alloy's expressions/formulas are analog to statements in imperative languages, except that they are hierarchical, i.e. expressions/formulas may contain other expressions/formulas.

AlloyFL locates faults at the AST node granularity and can locate any faulty expression, formula or paragraph (i.e. signature, predicate, function, fact and assertion) in an Alloy model. An Alloy test is typically invoked with an Alloy command (i.e. run or check) with an optional "expect" keyword, where expect 1 and expect 0 indicate satisfiability and unsatisfiability of the formula being invoked, respectively. If the invocation of a command for a test is satisfiable (or unsatisfiable) but the expected result is unsatisfiable (or satisfiable), then we say the test fails. The Alloy run commands invoke Alloy predicates or functions while the check commands invoke Alloy assertions. If a run command does not have the expect keyword, then we would expect its invocation to be satisfiable. On the contrary, we expect the invocation of a check command without the expect keyword to be unsatisfiable.

AUnit [55, 57] is a recent testing framework for Alloy and it provides the notion of test predicates (which are also ordinal Alloy predicates) that represent Alloy instances. MuAlloy [56, 61] is a recent mutation testing framework for Alloy that can automatically generate mutant killing AUnit test predicates. In this paper, since the availability of real faulty Alloy models (with manually written tests) is rather limited, we use Alloy run commands that invoke automatically generated test predicates to locate faults and evaluate AlloyFL. However, users can use AlloyFL with any test suite as long as some test fails.

AlloyFL contains 5 techniques: AlloyFL$_{co}$, AlloyFL$_{un}$, AlloyFL$_{su}$, AlloyFL$_{mu}$ and AlloyFL$_{hy}$. AlloyFL$_{co}$ implements the *spectrum-based FL* (SBFL) technique [4, 15, 20, 40] for Alloy. Since Alloy does not have control-flow and execution traces, all expressions/formulas in the same paragraph are either executed together or not executed at all. AlloyFL$_{co}$ statically analyzes Alloy paragraphs that are transitively used in each test. Then, AlloyFL$_{co}$ ranks the Alloy paragraphs based on the number of passing/failing tests that invoke the paragraphs and a suspiciousness formula. AlloyFL$_{un}$ implements a SAT-based technique, which leverages the unsat core [53, 58, 59]. AlloyFL$_{un}$ collects all AST nodes that are highlighted by the unsat core for each unsatisfiable failing test, and nodes highlighted more often are more likely to be faulty. AlloyFL$_{un}$ is designed to simulate how Alloy users would debug a faulty model manually using the unsat core. AlloyFL$_{su}$ is similar to AlloyFL$_{un}$ except that it uses both satisfiable and unsatisfiable failing tests to rank the AST nodes. AlloyFL$_{su}$ collects the nodes transitively used in satisfiable failing tests and nodes returned by unsat core in unsatisfiable failing tests. Nodes covered more often are ranked at the top. AlloyFL$_{mu}$ implements the *mutation-based FL* (MBFL) technique [39, 44] for Alloy. AlloyFL$_{mu}$ mutates Alloy AST nodes, e.g. "a&&b" to "a||b", to create non-equivalent mutants and check if the test results differ compared to the original model. AlloyFL$_{mu}$ uses a suspiciousness formula to compute the suspiciousness score for each mutant based on the number of passing/failing tests that kill the mutant. A test kills a mutant if its satisfiability changes compared to that of the original model. The node whose mutation gives the highest suspiciousness score, e.g. mutations on the node make almost all failing tests pass while preserving the results of passing tests, ranks at the top. The suspiciousness score of the mutated node conceptually propagate to all its descendants until mutations on its descendant nodes overwrite the corresponding suspiciousness scores. AlloyFL$_{hy}$ is a hybrid technique of both SBFL and MBFL. It takes the average of suspiciousness scores obtained from both AlloyFL$_{co}$ and AlloyFL$_{mu}$ and assign the score to the corresponding AST node. If an AST node is not mutable and does not have a suspiciousness score from AlloyFL$_{mu}$, then the suspiciousness score from AlloyFL$_{co}$ is used. Finally, if multiple nodes have the same suspiciousness score, then AlloyFL prioritizes the nodes with less descendants.

AlloyFL does not rank all AST nodes because Alloy does not have the notion of control flow and many AST nodes are equally suspicious. Additionally, previous studies have shown that users are unlikely to inspect more than a few candidates [23, 45]. To reduce the number of returned AST nodes, AlloyFL returns the root node of the AST subtree that contains the highest number of equally suspicious AST nodes compared to the root node. Many existing metrics, e.g. LIL [39], T-score [30], Expense [19], EXAM [64] and AWE [7], may not capture the proximity between the returned AST nodes and the faulty nodes. For example, AlloyFL may return a suspicious node that is the direct parent of a faulty node but the faulty node itself does not appear in the ranked list. In this case none of the above metrics reflect the closeness between the returned suspicious node and the faulty node. In this paper, we follow the spirit of the nearest neighbor distance metric (*NN*) in program dependence graphs (PDG) [48] to quantitatively measure the closeness between the ranked nodes and the faulty nodes. Specifically, we view the Alloy AST as PDG and adapt the *NN* distance metric on the AST. We design 3 distance metrics following *NN* and use the existing top-k metrics [65, 69], i.e. the number of faulty nodes in the top *k* returned nodes, to evaluate AlloyFL. The results on 38 real faults and 9000 mutant faults show that AlloyFL$_{mu}$ and AlloyFL$_{hy}$ are significantly more accurate than the baseline techniques, i.e. AlloyFL$_{co}$, AlloyFL$_{un}$ and AlloyFL$_{su}$.

This paper makes the following contributions:

- We propose, AlloyFL, the first set of AST node level FL techniques for Alloy that leverage multiple tests.
- We follow the spirit of an existing nearest neighbor distance metric [48] and define 3 new distance metrics at the AST level to measure the accuracy of AlloyFL.
- We evaluate AlloyFL using 38 real faults and 9000 mutant faults derived from 18 existing models. The subject models all contain 1 or more faults and our experimental results show that MBFL techniques are significantly more accurate than the baseline SBFL techniques and SAT-based techniques.
- We plan to release AlloyFL as well as 38 real faults and 9000 mutant faults so researchers can use and reproduce our results.

## 2 EXAMPLE MODEL

This section presents a real-world faulty Alloy model to introduce key concepts for AlloyFL. We briefly describe the basics of Alloy and AUnit as needed.

Figure 1a shows a faulty Alloy model of the well-known "farmer river-crossing" puzzle where the goal is to allow a farmer to transport a chicken, fox, and grain from one river bank to the other on a boat. However, the farmer can only carry one belonging on the boat at a time, and if left unattended, the fox will eat the chicken and the chicken will eat the grain. The model contains a modeling error which prevents the "eating" from happening while the farmer is away, and instead requires the farmer to be back. Figure 1b shows an AUnit test that fails.

The *signature* (sig) declaration, "sig Object", introduces a set of object atoms; abstract means the Object signature cannot have atoms of its own type, but its subsignatures can have atoms. eats is a *relation* that maps an Object atom to a set of Object atoms. Farmer, Fox, Chicken and Grain are declared as singleton subsignatures of Object. The *fact* eating states that the fox eats the chicken and the chicken eats the grain. Note that any *fact* in Alloy is enforced to be true. Signature State models the objects in both the near and far banks after every farmer's cross-river movement. The open declaration linearly orders the State atoms. The fact initialState constrains that initially everything is on the near bank and nothing is on the far bank. The *predicate* crossRiver defines the river crossing action. It takes four parameters (2 pairs of pre and post states): the set of objects on the bank where the farmer starts at (pre-state:from and post-state:from') and crosses to (pre-state:to and post-state:to') before and after the cross-river movement. The predicate states that either the farmer takes nothing or the farmer takes one item not including himself to the other side of the river. For the case when the farmer takes nothing the model uses a conjunction formula: "from' = from - Farmer && to' = to - to.eats + Farmer" which means that after the farmer crosses the river from bank, say *A* to bank, say *B*, the farmer is removed from the set of objects on bank *A* and added to the set of objects on bank *B*. When moving, the objects on bank *B* could change because an object may eat another object before the farmer's arrival. For the case when the farmer takes an item, the model uses an existentially quantified formula: "some item: from - Farmer | ...". The fact stateTransition states that for every two consecutive states, if the farmer is on the near bank in the pre-state, then he would cross the river to the far bank. Otherwise, he would cross the river from the far bank to the near bank. The predicate solvePuzzle restricts that in the last state, everything should be on the far bank.

The faults are in the predicate crossRiver and are colored in orange. The predicate considers eating to happen in "to" instead of "from", which stops the farmer from leaving and letting the fox eat

```
open util/ordering[State] as ord
abstract sig Object { eats: set Object }
one sig Farmer, Fox, Chicken, Grain extends Object {}
fact eating {
  eats = Fox->Chicken + Chicken->Grain  }
sig State { near,far: set Object }
fact initialState {
  let s0 = ord/first | s0.near = Object && no s0.far  }
pred crossRiver[from,from',to,to': set Object] {
  (from' = from - Farmer && to' = to - to.eats + Farmer )
  || (some item: from - Farmer {
  from' = from - Farmer - item && to' = to - to.eats + Farmer + item  }) }
fact stateTransition {
  all s: State, s': ord/next[s] {
    Farmer in s.near => crossRiver[s.near, s'.near, s.far, s'.far]
    else crossRiver[s.far, s'.far, s.near, s'.near]  }
}
pred solvePuzzle { ord/last.far = Object }
```

(a) Faulty Farmer Model in Alloy 4.1 release

```
pred test1 {
  some disj F0: Farmer | some disj X0: Fox |
  some disj C0: Chicken | some disj G0: Grain |
  some disj F0, X0, C0, G0: Object |
  some disj S0, S1, S2, S3: State {
    Farmer = F0
    Fox = X0
    Chicken = C0
    Grain = G0
    Object = F0 + X0 + C0 + G0
    eats = X0->C0 + C0->G0
    State = S0 + S1 + S2 + S3
    near = S0->F0 + S0->X0 + S0->C0 + S0->G0 + S1->X0 + S2->F0 + S2->X0 + S3->X0
    far = S1->F0 + S1->G0 + S2->G0 + S3->F0 + S3->G0
    ord/first = S0
    ord/next = S0->S1 + S1->S2 + S2->S3
    crossRiver[F0 + X0 + C0 + G0, C0, none, F0 + X0] } }
run test1 for 4 expect 1
// More tests ...
```

(b) MuAlloy Generated Tests

Figure 1: Faulty Farmer Example and MuAlloy Generated Tests.

the chicken without the farmer coming back. The correct formula should be: "from' = from - Farmer - from'.eats && to' = to + Farmer" and "from' = from - Farmer - item - from'.eats && to' = to + Farmer + item". This modeling error is introduced in Alloy release 4.1 and fixed in release 4.2. An automatically generated AUnit test that reveals the fault is shown in Figure 1b. Predicate test1 encodes the valuation of each signature type in the farmer model in Figure 1a. Each relation is assigned some atoms, e.g. Farmer contains a single atom F0. The invocation of crossRiver predicate states that all objects are on the near bank in the pre-state and nothing (none) is on the far bank. In the post-state (after the farmer crosses the river with the fox), only the chicken is left on the near bank (because the chicken is supposed to eat the grain) and both the farmer and the fox are on the far bank. The command "run test for 4 expect 1" runs the test with a scope of at most 4 atoms for each signature type and expects the existence of a solution. However, the faulty farmer model does not have any solution with respect to this test, which contradicts the expectation and causes a test failure.

We use a test suite that contains some failing tests (e.g. test1) to locate the fault using AlloyFL. AlloyFL$_{co}$ assigns all Alloy paragraphs equal suspiciousness scores (except the fact solvePuzzle as it is never covered by any failing tests) because all tests implicitly invoke facts and the stateTransition fact invokes the crossRiver predicate. The most suspicious AST nodes are highlighted in red (including yellow and green) in Figure 1a. Both AlloyFL$_{un}$ and AlloyFL$_{su}$ report the entire body of the crossRiver predicate as the most suspicious AST node which is highlighted in yellow (including green) . AlloyFL$_{mu}$ and AlloyFL$_{hy}$ report the node "from' = from - Farmer - item && to' = to - to.eats + Farmer + item" as the most suspicious node because mutating the root node && makes the most failing tests pass compared to mutating other AST nodes. Thus, the most suspicious node && and its descendants are highlighted in green. We can see that the most suspicious node returned by AlloyFL$_{mu}$ and AlloyFL$_{hy}$ is closest to the faulty node. Thus, the MBFL techniques are most accurate among all techniques in this example.

| Name | Formula |
|---|---|
| Tarantula [19] | $\frac{\frac{failed(e)}{totalfailed}}{\frac{failed(e)}{totalfailed} + \frac{passed(e)}{totalpassed}}$ |
| Ochiai [2] | $\frac{failed(e)}{\sqrt{totalfailed \times (failed(e) + passed(e))}}$ |
| Op2 [40] | $failed(e) - \frac{passed(e)}{totalpassed + 1}$ |
| Barinel [3] | $1 - \frac{passed(e)}{passed(e) + failed(e)}$ |
| DStar [63] | $\frac{failed(e)^*}{passed(e) + (totalfailed - failed(e))}$ |

*totalfailed*: the total number of test cases that failed.
*totalpassed*: the total number of test cases that pass.
*failed(e)*: the number of failed test cases that cover or kill *e*.
*passed(e)*: the number of passed test cases that cover or kill *e*.

Figure 2: Suspiciousness Formulas in AlloyFL.

## 3 TECHNIQUE

In this section, we describe the formulas to compute suspiciousness scores (Section 3.1) and all techniques in AlloyFL (Section 3.2).

### 3.1 Suspiciousness Formulas

Figure 2 shows the formulas that AlloyFL$_{co}$, AlloyFL$_{mu}$ and AlloyFL$_{hy}$ support to compute the suspiciousness score. For AlloyFL$_{co}$, the code elements (*e*) are AST nodes. For AlloyFL$_{mu}$ and AlloyFL$_{hy}$, killed mutants are treated as covered code elements while live mutants are treated as uncovered code elements. *totalfailed* and *totalpassed* are the number of test cases which failed and passed w.r.t. the original model. *failed(e)* and *passed(e)* are the number of test cases which failed and passed that cover the AST node or kill the mutant *e*.

### 3.2 AlloyFL

AlloyFL locates faults at the AST node granularity, which allows it to locate faulty expressions or formulas that are hierarchical. We present AlloyFL$_{co}$ as the first baseline technique and expect it to be inaccurate because of the non-existence of control flow in the Alloy language. We next present two baseline techniques, i.e. AlloyFL$_{un}$ and AlloyFL$_{su}$ which simulate what Alloy users can achieve by using the unsat core. Finally, we present two more advanced techniques AlloyFL$_{mu}$ and AlloyFL$_{hy}$ which define a diverse set of mutation operators and are shown to be more accurate.

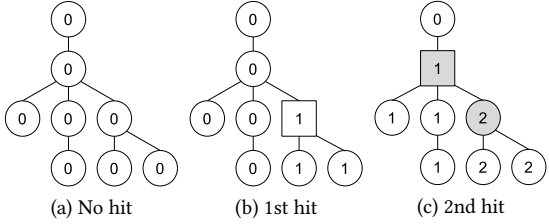

Figure 3: Illustration of AlloyFL$_{un}$ and AlloyFL$_{su}$

**AlloyFL$_{co}$.** Since Alloy does not have control-flow and execution traces, every code element in the same paragraph will be either executed together or not executed at all by a given test. This means nodes declared in the same paragraph would share the same suspiciousness score. To implement AlloyFL$_{co}$, we build a static analyzer which analyzes the entire AST and binds a variable usage or a predicate/function call to its signature or predicate/function declaration. The static analyzer is used to find all Alloy paragraphs transitively used by a test. However, the analyzer ignores dependencies that are never used. For example, if a test uses a formula "all s: S, t: T | some s && p[s]" where variable "t" is not used and "p[s]" is a predicate invocation, then the test only depends on signature "S" and predicate "p". By default, all facts are implicitly used, and all paragraphs transitively invoked in the facts and the test predicate are covered by the test. AlloyFL$_{co}$ computes a suspiciousness score for each Alloy paragraph based on the number of passing/failing tests that cover it and a formula shown in Figure 2. Finally, all paragraphs are ranked in descending order of suspiciousness score. In case of a tie, the paragraph which has a smaller number of descendants is prioritized.

**AlloyFL$_{un}$.** To implement AlloyFL$_{un}$, we modify the standard Alloy toolset to return AST nodes when the MiniSat solver with unsat core is used [58, 59]. We configure the solver such that it is guaranteed to return a local minimum core and all formulas are fully expanded (pushing negations in as much as possible, removing existential quantifiers using skolemization and expanding universal quantifiers given the bounds on the signatures) to make the returned core as fine-grained as possible. AlloyFL$_{un}$ constructs a hit-map for the entire AST and every node in the AST has a count initially set to 0. If a node is returned by the unsat core, then the count for the node itself *and* each of its descendant increases by 1. To illustrate, Figure 3 shows how the hit-map is built. Initially, each node has a count of 0 (Figure 3(a)). In Figure 3(b), a node denoted by the square is returned by the unsat core and AlloyFL$_{un}$ increases the counts of all the affected descendants. This process applies for all the subsequently returned nodes. For example, suppose the square node in Figure 3(c) is returned next, the count of each descendant is increased to 1 and the count of each previously hit node is increased to 2. Note that a child node always has a count greater than or equal to its parent's count. AlloyFL$_{un}$ collects every node whose count is greater than its parent's count, e.g. the gray

**Algorithm 1:** Sat-Unsat Based Fault Localization

**Input**: Faulty Alloy model *M*, test suite *T*.
**Output**: Ranked list of suspicious AST nodes *L*.

$L \leftarrow []$, $L' \leftarrow []$, $R$ = runTests(*M*, *T*)
**foreach** $r \in R$ **do**
    **if** *r.isPassed()* **then continue**
    **if** *r.isSatisfiable()* **then** *L'*.add(staticAnalyze(*r*))
    **else** *L'*.add(unsatCore(*r*))
*hitmap* ← <Node,Int>{} // Default value is 0
**foreach** *nodes* ∈ *L'* **do**
    **foreach** *n* ∈ *nodes* **do**
        **foreach** *d* ∈ *n.getDesc()* **do** *hitmap*[*d*] += 1
**foreach** *n* ∈ *hitmap* **do**
    **if** *hitmap*[*n.getParent()*] < *hitmap*[*n*] **then** *L*.add(*n*)
*L*.sortByHitAndSize(*hitmap*, reverse=**True**)
**return** *L*

nodes in Figure 3(c). AlloyFL$_{un}$ does not collect the root node as we set the root's parent to null. The collected nodes are ranked in descending order of the corresponding count. In case of a tie, nodes with a smaller number of descendants are prioritized. Note that AlloyFL$_{un}$ only works for unsatisfiable tests and cannot be used if the model is strictly underconstrained, in which case no unsatisfiable failing test exists.

**AlloyFL$_{su}$.** Similar to AlloyFL$_{un}$, AlloyFL$_{su}$ also constructs a hitmap for the entire AST. The difference is that AlloyFL$_{su}$ uses nodes reported from both unsatisfible and satisfiable failing tests. The nodes reported from the unsatisfiable failing tests are the same as for AlloyFL$_{un}$, and the nodes reported from the satisfiable failing tests are from the static analyzer described for AlloyFL$_{co}$. We give the official algorithm for AlloyFL$_{su}$ in Algorithm 1. The algorithm takes as input a faulty model *M* and a test suite *T*, and returns the ranked list of the suspicious AST nodes *L*. *L'* keeps the nodes returned by the static analyzer and the unsat core. Both *L* and *L'* are initialized to empty lists. The algorithm collects the test results *R* by invoking *T* over *M*. For each individual test result *r*, we skip if *r* is passed. If *r* fails and is satisfiable, then we collect all transitively used nodes of the corresponding test by invoking the static analyzer and add those nodes to *L'*. If *r* fails and is unsatisfiable, we collect all nodes returned by the unsat core and add them to *L'*. Note that *L'* is a list of sets of nodes. To sort the nodes, we first initialize a *hitmap* as an empty map with a default value of 0. For every set of *nodes* in *L'*, we increase the counts of each individual node *n* in *nodes* and *n*'s descendants in the *hitmap*. Then, for each node *n* whose count is bigger than its parent's count, we add it to *L*. Finally, we sort *L* in descending order of the number of times a node is hit and prioritize nodes with a smaller number of descendants in case of a tie. Algorithm 1 boils down to AlloyFL$_{un}$ if we do not collect nodes when the test is satisfiable. The intuition of the algorithm is that nodes covered by more failing tests are more likely to be faulty, and we use nodes returned by the unsat core if possible because the core typically gives finer grained nodes compared to the static analyzer.

| Mutation Operator | Description |
|---|---|
| MOR | Multiplicity Operator Replacement |
| QOR | Quantifier Operator Replacement |
| UOR | Unary Operator Replacement |
| BOR | Binary Operator Replacement |
| LOR | Formula List Operator Replacement |
| UOI | Unary Operator Insertion |
| UOD | Unary Operator Deletion |
| LOD | Logical Operand Deletion |
| PBD | Paragraph Body Deletion |
| BOE | Binary Operand Exchange |
| IEOE | Imply-Else Operand Exchange |

**Figure 4: Mutation Operators.**

***AlloyFL$_{mu}$***. AlloyFL$_{mu}$ implements a wide variety of mutation operators as shown in Figure 4. *MOR* mutates signature multiplicity, e.g. "one sig" to "lone sig". *QOR* mutates quantifiers, e.g. some to all. *UOR*, *BOR* and *LOR* define operator replacement for unary, binary and formula list operators, respectively. For example, *UOR* mutates a.^b to a.*b; *BOR* mutates a<=>b to a=>b; and *LOR* mutates a||b to a&&b. *UOI* inserts an unary operator before expressions, e.g. a.b to a.~b. *UOD* deletes an unary operator, e.g. a.^~b to a.^b. *LOD* deletes an operand of a logical operator, e.g. a&&b to b. *PBD* deletes the body of an Alloy paragraph. *BOE* exchanges operands for a binary operator, e.g. a-b to b-a. *IEOE* exchanges the operands of imply-else operation, e.g. "a => b else c" to "a => c else b".

Algorithm 2 shows the details of AlloyFL$_{mu}$. The algorithm takes as input a faulty Alloy model *M*, a test suite *T*, a set of mutation operators *Ops* and a suspiciousness formula *F*. The output of the algorithm is a ranked list of suspicious AST nodes (*L*) sorted in the descending order of suspiciousness. Initially, *L* is set to an empty list. *S* keeps the set of nodes covered by failing tests and is initialized as an empty set. AlloyFL$_{mu}$ runs *T* against *M* and the results are stored in *R*. *n2s* keeps the mapping from a node to its suspiciousness score and it is initialized to an empty map with a default value of 0. For each test result *r* in *R*, AlloyFL$_{mu}$ collects nodes and their descendants covered by all failing tests. Then, AlloyFL$_{mu}$ iterates over each node *n* in *M*. If *n* is not covered by any failing test, i.e. $n \notin S$, then AlloyFL$_{mu}$ skips it. For each *n* covered by the failing tests, AlloyFL$_{mu}$ tries to apply every mutation operator in *Ops* to the node, one at a time. If the mutation operator is not applicable, it is skipped. Otherwise, AlloyFL$_{mu}$ mutates *M* to *M'*. If *M'* leads to a compilation error or is equivalent to *M*, then AlloyFL$_{mu}$ skips *M'*. Otherwise, AlloyFL$_{mu}$ runs *T* against the mutant *M'* and collects the result as *R'*. Function *computeSusp* computes the suspiciousness score of the mutant based on the formula *F* (Figure 2), and test results *R* and *R'*. *n2s* keeps the maximum suspiciousness score for each node *n*. After AlloyFL$_{mu}$ exhausts all mutation operators that are applicable to *n*, *n* is added to *L* if its suspiciousness score *n2s[n]* is greater than 0. Finally, after all AST nodes are exhausted, *L* is sorted in descending order of suspiciousness and returned.

***AlloyFL$_{hy}$***. Inspired by [47], AlloyFL$_{hy}$ assigns the average of suspiciousness scores calculated from both AlloyFL$_{co}$ and AlloyFL$_{mu}$ to each AST node. If a node is not mutable, then AlloyFL$_{hy}$ uses the same suspiciousness score as AlloyFL$_{co}$. The intuition of AlloyFL$_{hy}$ is that AlloyFL$_{mu}$ sometimes perform badly for omission errors in which case AlloyFL$_{co}$ performs relatively well. So AlloyFL$_{hy}$ is designed to combine the strengths of both AlloyFL$_{co}$ and AlloyFL$_{mu}$.

**Algorithm 2:** Mutation-Based Fault Localization

**Input**: Faulty Alloy model *M*, test suite *T*, mutation operators *Ops*, suspiciousness formula *F*.
**Output**: Ranked list of suspicious AST nodes *L*.

$L \leftarrow []$, $S \leftarrow \emptyset$, $R$ = runTests(*M*, *T*)
*n2s* ← <Node, Double>{} // Default value is 0.0
**foreach** $r \in R$ **do**
    **if** *r.isPassed()* **then continue**
    **foreach** $n \in staticAnalyze(r)$ **do** *S*.addAll(*n*.getDesc())
**foreach** $n \in M.getNodes()$ **do**
    **if** $n \notin S$ **then continue**
    **foreach** $op \in Ops$ **do**
        **if** *!isApplicable(op, n)* **then continue**
        *M'* = applyOp(*op, n, M*)
        **if** *isValid(M')* && *!isEquivalent(M, M')* **then**
            *R'* = runTests(*M', T*)
            *n2s*[*n*] = max(*n2s*[*n*], computeSusp(*F, R, R'*))
    **if** *n2s*[*n*] > 0 **then** *L*.add(*n*)
*L*.sortByScore(*n2s*, reverse=**True**)
**return** *L*

## 4 DISTANCE METRICS

To quantitatively measure how close the ranked nodes are to the real faulty nodes, we follow the spirit of the nearest neighbor distance metric (*NN*) in the program dependence graph (PDG) [48]. Since there is no notion of control dependences in declarative languages like Alloy, we view the Alloy AST as a PDG and adapt the *NN* distance metric on the AST.

The original nearest neighbor distance metric quantifies the percentage of nodes not needing inspection by the programmer using the formula $1 - \frac{|S(R)|}{|G|}$, where $R = \{n_1, n_2, ..., n_k\}$ is the top k returned suspicious nodes ($n_i, 1 \leq i \leq k$), $S(R)$ is a sphere of all nodes in the graph *G* such that the maximum distance of any node in *S* to its closest suspicious node is smaller or equal to the minimum distance of any suspicious node in *R* to its closest faulty node. Conceptually, the user does a breadth-first search starting with the suspicious nodes, and increasing the distance until a defect is found. The formula computes the percentage of nodes that need not be examined. However, previous studies show that: (1) the percentage of nodes needing inspection is a better estimate than the percentage of nodes not needing inspection [30, 64]; and (2) fault localization techniques should focus on improving absolute rank rather than percentage rank [45]. Therefore, we enhance the *NN* metric to use the absolute number of nodes needing inspection ($|S(R)|$). Techniques which give smaller distance metric values are more accurate. We next describe 3 distance metrics used to evaluate AlloyFL.

***Nearest Neighbor Up-Down (NNUD)***. NNUD sets *R* to the *k* most suspicious nodes returned. It allows traversing upward (parent) and downward (children) from the suspicious nodes in the AST until a faulty node is found. Figure 5(a) shows the number of nodes one needs to explore from the top two suspicious nodes. The number in the circle represents the position of the node in the ranked list, e.g. 1 means it ranks at the top. "F" shows the faulty node and squares are irrelevant nodes. Circles colored in gray estimate the nodes users need to inspect under NNUD metric with $k = 2$. Since the minimum distance between any of the two suspicious nodes and the faulty node is 1, all nodes that are reachable from the suspicious nodes within a distance of 1 are included. Thus, the metric reports 6, i.e. the size of the gray nodes. NNUD assumes that the programmer may look at the parent or children when inspecting the top $k$ suspicious nodes until a faulty node is found.

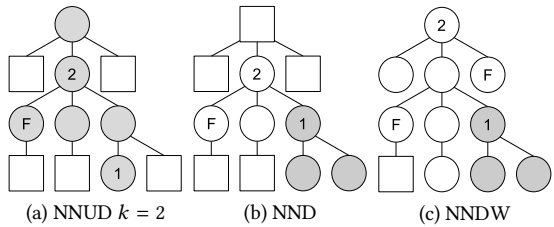

(a) NNUD $k = 2$  (b) NND  (c) NNDW

**Figure 5: Distance Metrics Examples**

*Nearest Neighbor Down (NND)*. NND does not allow traversing upward from the suspicious node and it processes suspicious nodes one at a time. Figure 5(b) shows how the metric works. From the top 1 suspicious node, we can only traversing downward. Since no faulty node is found, we mark all inspected nodes in gray. Then, NND does a breadth-first search for the second top suspicious node. In this case, a faulty node is found and all descendants within the same distance, i.e. $\leqslant 1$, are included (3 circles colored in white), excluding already visited nodes colored in gray. Finally, NND reports 6, i.e. the size of the inspected nodes in circle. This metric assumes that the users only inspect the children and never reinspect already visited nodes. However, it is possible that the faulty nodes never appear as the descendants of any suspicious node. To avoid this scenario, we append the root node of the entire AST to the end of the ranked suspicious node list returned by AlloyFL. This makes sure that the metric always terminates with a faulty node found.

*Nearest Neighbor Down Worst (NNDW)*. NNDW is similar to NND (only allows traversing downward) except that it assumes the user is unlucky and would inspect all non-faulty nodes before finding the fault. Figure 5(c) shows how the metric works. Inspecting the top suspicious node is similar to NND, with the difference occurring when inspecting the second top suspicious node. In this case, we traverse downward and include all non-faulty nodes that have not been visited before (white circles without the faulty node). If a faulty node can be reached from the current suspicious node, then we stop traversing and include all such faulty nodes. In this case, two faulty nodes appear as the children of the second top suspicious node, so we include both faulty nodes. Finally, NNDW returns 10, i.e. all circle nodes. Similar to NND, we append the root node of the entire AST to the end of the suspicious node list.

## 5 EVALUATION

We evaluate AlloyFL on 38 real faults collected from Alloy release 4.1, Amalgam [42] and graduate student solutions. These faulty models contain various types of faults, including overconstraints, underconstraints and a mixture of both. In addition, We also extend MuAlloy with the ability to generate higher order mutants [18] and evaluate AlloyFL on 9000 mutant models with exactly two mutant faults. All experiments are performed on Ubuntu 16.04 LTS with 2.4GHz Intel Xeon CPU and 8 GB memory.

In this section, we address the following research questions for both real faults and mutant faults:

- **RQ1.** What is the accuracy and time overhead of AlloyFL?
- **RQ2.** How does the suspiciousness formula affect AlloyFL?

### 5.1 Experiment Setting

Table 6 gives an overview for the 18 correct models used to generate mutant faults in the evaluation. Address book (**addr**), Dijkstra mutex algorithm (**dijkstra**), farmer cross-river puzzle (**farmer**), and Halmos handshake problem (**hshake**) are from Alloy's example set. Bad employee (**bempl**), grade book (**grade**), and other groups (**other**) are Alloy translations of access-control specifications used to benchmark Amalgam [42]. Binary tree (**bt**), colored tree (**ctree**), full tree (**fullTree**), n-queens problem (**nqueens**) and singly-linked list (**sll**) are from MuAlloy [61]. Array model (**array**), balanced binary search tree (**bst**), class diagram (**cd**), doubly-linked list (**dll**), finite state machine (**fsm**), and singly-linked list with sorting and counting functions (**stu**) are homework questions we assigned to graduate students.

For each subject, Figure 6 shows the number of AST nodes (**ast**), the number of nonequivalent first-order mutants (**1st**), the number of tests *automatically generated* (**tot**), the number of tests that are expected to be satisfiable (**sat**) and unsatisfiable (**uns**), the number of nonequivalent second-order mutants (**2nd**), and the scope used to run tests or equivalence checks (**scp**). Prior works shows that test cases generated by MuAlloy are effective

| Model | ast | 1st | test | | | 2nd | scp |
|---|---|---|---|---|---|---|---|
| | | | tot | sat | uns | | |
| addr | 124 | 62 | 30 | 19 | 11 | 2.0k | 3 |
| array | 68 | 51 | 36 | 14 | 22 | 1.3k | 3 |
| bst | 175 | 167 | 110 | 50 | 60 | 21.1k | 4 |
| bempl | 57 | 35 | 25 | 11 | 14 | 594 | 3 |
| bt | 61 | 74 | 34 | 20 | 14 | 3.4k | 3 |
| cd | 52 | 46 | 24 | 9 | 15 | 1.6k | 3 |
| ctree | 76 | 83 | 22 | 9 | 13 | 4.8k | 3 |
| dijkstra | 410 | 183 | 120 | 44 | 76 | 19.7k | 3 |
| dll | 92 | 81 | 48 | 22 | 26 | 3.6k | 3 |
| farmer | 180 | 106 | 56 | 33 | 23 | 6.5k | 4 |
| fsm | 85 | 63 | 15 | 3 | 12 | 3.0k | 3 |
| fullTree | 85 | 100 | 44 | 24 | 20 | 6.5k | 3 |
| grade | 77 | 44 | 41 | 23 | 18 | 978 | 3 |
| hshake | 136 | 107 | 33 | 10 | 23 | 10.3k | 4 |
| nqueens | 110 | 101 | 75 | 36 | 39 | 5.3k | 4 |
| other | 40 | 32 | 21 | 9 | 12 | 558 | 3 |
| sll | 38 | 31 | 22 | 14 | 8 | 574 | 3 |
| stu | 201 | 143 | 87 | 40 | 47 | 14.2k | 3 |
| **Sum** | 2.1k | 1.5k | 843 | 390 | 453 | 106.0k | |

**Figure 6: Correct Models Information.**

in detecting real faults [56, 61], so we use MuAlloy to generate first-order non-equivalent mutants and the corresponding tests that kill the mutants. We choose to generate second-order mutants for the injected faults because (1) it quickly becomes time consuming to generate mutants with an order higher than 2; and (2) we want to enhance the credibility of our results by using models with more than 1 fault. We filter out second-order mutants that cannot be killed by the generated test suite to make sure at least one fault can be revealed by the test suite. For real faults, we manually inspect all of the faults and try to fix them without changing the model structure. For example, if the model has a fault in the

| Model | nnud1 ||||| nnud5 ||||| nnud10 ||||| nnd ||||| nndw |||||
|---|---|---|---|---|---|---|---|---|---|---|---|---|---|---|---|---|---|---|---|---|---|---|---|---|---|
| | Co | Un | Su | Mu | Hy | Co | Un | Su | Mu | Hy | Co | Un | Su | Mu | Hy | Co | Un | Su | Mu | Hy | Co | Un | Su | Mu | Hy |
| addr1 | 78 | 24 | 24 | **1** | **1** | 78 | 57 | 48 | **5** | **5** | 73 | 57 | 61 | **10** | **10** | 50 | 10 | 10 | **1** | **1** | 57 | 15 | 15 | **1** | **1** |
| arr1 | 51 | 20 | 20 | **14** | **14** | 51 | 56 | 56 | 38 | **34** | 51 | 61 | 61 | 38 | 49 | 32 | 61 | 61 | **8** | **8** | 41 | 64 | 64 | **15** | **15** |
| arr2 | 29 | **1** | **1** | **1** | **1** | 26 | **2** | **2** | 4 | 5 | 26 | **2** | **2** | 4 | 8 | 22 | **1** | **1** | **1** | **1** | 52 | **3** | **3** | **3** | **3** |
| bst1 | 32 | 32 | 32 | **5** | **5** | 58 | 55 | 55 | **5** | **5** | 65 | 55 | 55 | **6** | 10 | **5** | **5** | **5** | **5** | **5** | **5** | **5** | **5** | **5** | **5** |
| bst2 | 3 | 3 | 40 | 4 | **1** | 15 | 15 | 53 | 14 | **5** | 28 | 15 | 23 | **10** | **10** | **2** | **2** | 64 | 7 | **1** | 6 | 6 | 76 | 7 | **5** |
| bst3 | 19 | 68 | 68 | **1** | **1** | 47 | 34 | 32 | **5** | **5** | 62 | 34 | 40 | **10** | **10** | 6 | 14 | 14 | **1** | **1** | 9 | 17 | 17 | **1** | **1** |
| bempl1 | 19 | 19 | 19 | **3** | **3** | **7** | 19 | **7** | 10 | 10 | **17** | 19 | 18 | **17** | 19 | 6 | 19 | 6 | **2** | **2** | 6 | 48 | 6 | **2** | **2** |
| cd1 | 23 | 45 | 45 | **4** | **4** | 38 | 45 | 40 | **5** | **5** | 45 | 45 | 45 | **10** | **10** | 9 | 45 | 15 | **3** | **3** | 14 | 58 | 21 | **5** | **5** |
| cd2 | 16 | 34 | 34 | **1** | **1** | 25 | 34 | 31 | **4** | 5 | 25 | 34 | 31 | **4** | 6 | 6 | 34 | 12 | **1** | **1** | 9 | 47 | 16 | **2** | **2** |
| ctree1 | **18** | **18** | **18** | 39 | 39 | **5** | 18 | 9 | 56 | 56 | **8** | 18 | 17 | 21 | 21 | **4** | 18 | 5 | 61 | 47 | **4** | 67 | 5 | 71 | 51 |
| dll1 | 6 | **4** | **4** | **4** | **4** | 30 | 16 | 16 | 19 | **14** | 33 | **16** | **16** | 19 | 26 | **4** | **3** | **3** | **3** | **3** | 7 | **6** | **6** | **6** | **6** |
| dll2 | 6 | 4 | 4 | **1** | **1** | 30 | 16 | 14 | **5** | **5** | 36 | 16 | 22 | **10** | **10** | 4 | 3 | 3 | **1** | **1** | 8 | 7 | 7 | **1** | **1** |
| dll3 | **1** | 31 | 31 | **2** | **2** | **5** | 26 | 38 | **5** | **5** | **7** | 26 | 22 | **10** | **10** | **1** | 11 | 11 | **2** | **2** | **1** | 13 | 13 | **2** | **2** |
| dll4 | 6 | 4 | 4 | **1** | **1** | 30 | 16 | 16 | 4 | **5** | 33 | 16 | 16 | **4** | 8 | 4 | 3 | 3 | **1** | **1** | 8 | 7 | 7 | **1** | **1** |
| farmer1 | 90 | **17** | **17** | **9** | **9** | 90 | 94 | 94 | **26** | **26** | 79 | 94 | 94 | **40** | 43 | 54 | 13 | 13 | **7** | **7** | 74 | 30 | 30 | **12** | **12** |
| fsm1 | 50 | 4 | 4 | **1** | **1** | 49 | 19 | 15 | **5** | **5** | 50 | 19 | 25 | **10** | **10** | 71 | 3 | 3 | **1** | **1** | 91 | 7 | 7 | **1** | **1** |
| fsm2 | 59 | 59 | 59 | **1** | **1** | 59 | 59 | 59 | **5** | **5** | 59 | 59 | 59 | **10** | **10** | 69 | 59 | 44 | **1** | **1** | 78 | 87 | 52 | **1** | **1** |
| grade1 | **1** | 19 | 19 | 23 | **4** | **5** | 19 | 27 | 15 | **5** | **7** | 19 | 18 | 25 | **10** | **1** | 19 | 21 | 42 | 18 | **1** | 56 | 29 | 71 | 18 |
| other1 | 27 | 27 | 27 | 28 | 28 | 28 | 27 | 28 | **5** | **5** | 17 | 27 | 23 | **7** | 10 | 14 | 27 | 15 | **8** | **8** | 18 | 60 | 19 | **9** | **9** |
| stu1 | 7 | **4** | **4** | 11 | 11 | 21 | 18 | 18 | **8** | 9 | 41 | **18** | **18** | 24 | 19 | 4 | **3** | **3** | 5 | 5 | 12 | **11** | **11** | **11** | **11** |
| stu2 | 9 | **4** | **4** | **4** | **4** | 26 | 18 | 10 | **5** | **5** | 47 | 18 | 26 | **10** | **10** | **4** | **3** | **3** | 4 | 4 | 6 | 11 | 11 | **5** | **5** |
| stu3 | 3 | **1** | **1** | 3 | 3 | 9 | **4** | 5 | 13 | 11 | 24 | **4** | 10 | 10 | 10 | 2 | **1** | **1** | 2 | 2 | 10 | **4** | **4** | 10 | 10 |
| stu4 | 7 | **4** | **4** | 11 | 11 | 21 | 18 | 18 | **8** | 9 | 42 | **18** | **18** | 24 | 19 | 4 | **3** | **3** | 5 | 5 | 12 | **11** | **11** | **11** | **11** |
| stu5 | 7 | **4** | **4** | 11 | 11 | 21 | 18 | 18 | **8** | 9 | 42 | **18** | **18** | 24 | 19 | 4 | **3** | **3** | 5 | 5 | 12 | **11** | **11** | **11** | **11** |
| stu6 | 32 | 10 | 10 | **1** | **1** | 38 | 45 | 45 | **5** | **5** | 9 | 45 | 45 | **7** | 10 | 35 | 5 | 5 | **1** | **1** | 51 | 7 | 7 | **1** | **1** |
| stu7 | 3 | **1** | **1** | **1** | **1** | 9 | **5** | **5** | **5** | **5** | 24 | **5** | 10 | 10 | 10 | 2 | **1** | **1** | **1** | **1** | 10 | **7** | **7** | **7** | **7** |
| stu8 | 7 | **4** | **4** | **4** | **4** | 22 | 19 | 19 | 23 | **9** | 47 | **19** | **19** | 23 | 21 | 4 | **3** | **3** | **3** | **3** | 7 | **6** | **6** | **6** | **6** |
| stu9 | 9 | 66 | 66 | **1** | **1** | 36 | 66 | 66 | **5** | **5** | 47 | 66 | 60 | **10** | **10** | 4 | 66 | 86 | **1** | **1** | 5 | 166 | 98 | **1** | **1** |
| stu10 | 7 | **4** | **4** | 11 | 11 | 22 | 19 | 19 | **8** | 9 | 47 | **19** | **19** | 25 | 19 | **4** | **3** | **3** | 5 | 5 | 12 | **11** | **11** | **11** | **11** |
| stu11 | 23 | **8** | **8** | 11 | 11 | 41 | 38 | 38 | **10** | **10** | 70 | 38 | 38 | 27 | **19** | 7 | **6** | **6** | 9 | 9 | 17 | 16 | 16 | **15** | **15** |
| stu12 | **7** | 90 | 90 | 11 | 11 | 31 | 29 | 17 | **5** | **5** | 46 | 29 | 27 | **10** | **10** | **4** | 11 | 11 | **3** | **3** | 13 | 20 | 20 | **3** | **3** |
| stu13 | 45 | 116 | 116 | **1** | **1** | 63 | 116 | 116 | **5** | **5** | 98 | 116 | 102 | **10** | **10** | 11 | 116 | 28 | **1** | **1** | 18 | 204 | 39 | **1** | **1** |
| stu14 | 34 | 9 | 9 | **1** | **1** | 40 | 46 | 46 | **5** | **5** | 10 | 46 | 46 | **7** | 10 | 40 | 5 | 5 | **1** | **1** | 52 | 7 | 7 | **1** | **1** |
| stu15 | 7 | **4** | **4** | 11 | 11 | 21 | 22 | 22 | **8** | 9 | 43 | 22 | 22 | 24 | **19** | **4** | **3** | **3** | 5 | 5 | 12 | **11** | **11** | **11** | **11** |
| stu16 | **1** | 32 | 32 | 60 | **1** | **5** | 32 | 38 | 56 | **5** | **10** | 32 | 25 | 64 | **10** | **1** | 32 | 47 | 63 | **1** | **1** | 94 | 55 | 94 | **1** |
| stu17 | 32 | **4** | **4** | 11 | 11 | 38 | 22 | 22 | 11 | **9** | **9** | 22 | 22 | 9 | 19 | 55 | **3** | **3** | 5 | 5 | 69 | **11** | **11** | **11** | **11** |
| stu18 | 74 | 17 | 17 | **4** | **4** | 53 | **4** | 5 | 5 | 5 | 76 | **4** | 10 | 10 | 10 | 12 | **8** | **8** | 9 | 9 | 14 | **10** | **10** | **9** | **9** |
| stu19 | 3 | **1** | **1** | **1** | **1** | 12 | **5** | **5** | **5** | **5** | 24 | **9** | 10 | 10 | 10 | 2 | **1** | **1** | **1** | **1** | 4 | **1** | **1** | 3 | 3 |
| **Avg** | 22.4 | 21.5 | 22.4 | 8.2 | **6.1** | 31.7 | 30.8 | 30.8 | 11.4 | **9.1** | 38.9 | 31.1 | 31.4 | 15.9 | **14.6** | 14.9 | 16.5 | 14.0 | 7.5 | **4.7** | 21.7 | 32.2 | 19.6 | 11.5 | **7.1** |
| **Med** | 12.5 | 9.5 | 13.5 | 4.0 | **3.5** | 29.0 | 20.5 | 22.0 | 5.0 | **5.0** | 41.5 | 22.5 | 22.5 | 10.0 | **10.0** | 4.0 | 5.0 | 5.0 | 3.0 | **2.5** | 12.0 | 11.0 | 11.0 | 5.5 | **5.0** |
| **Std** | 23.0 | 26.4 | 26.4 | 11.8 | **7.8** | 20.2 | 24.1 | 24.6 | 12.6 | **9.6** | 22.7 | 24.2 | 22.2 | 11.9 | **8.9** | 19.9 | 23.9 | 19.7 | 14.5 | **7.8** | 24.4 | 43.9 | 21.6 | 20.4 | **8.7** |
| **Win** | 6 | 16 | 16 | 23 | **25** | 5 | 5 | 4 | 26 | **26** | 6 | 11 | 7 | 21 | **15** | 5 | 16 | 16 | 23 | **25** | 5 | 13 | 13 | 31 | **33** |

Figure 7: Distance Metrics for Real Faults.

quantifier body, then we try to fix it without replacing the entire quantifier formula. The expressions/formulas modified due to the fix are labeled as faulty. For mutant faults, following standard practice [24, 25, 66] the mutated nodes are labeled as faulty. We collect 5 real faults from [42], 1 real fault from Alloy release 4.1 and 32 real faults from graduate students. Additionally, we randomly sample 500 second-order mutants for each subject (9000 in total). The models we used to generate mutants contain all correct versions of the real faults.

To evaluate AlloyFL, we use both distance metrics, i.e. NNUD top1 (nnud1), NNUD top5 (nnud5), NNUD top10 (nnud10), NND(nnd) and NNDW(nndw), and the traditional top-k metric, i.e. number of faults in top1, top5 and top10 suspicious nodes. We pick $k$ up to 10 because [23] showed that 98% of practitioners consider a fault localization technique to be useful only if the fault appears in the top-10 suspicious elements. Techniques with smaller values of distance metrics and larger values of top-k metrics are more accurate.

### 5.2 RQ1: AlloyFL Accuracy and Time Overhead

Figure 7 shows the distance metric results, i.e. the number of AST nodes to inspect before finding the first fault, of AlloyFL for real faults. We use Ochiai formula for AlloyFL$_{co}$, AlloyFL$_{mu}$ and AlloyFL$_{hy}$. The most accurate AlloyFL techniques w.r.t. each distance metric are highlighted in bold. For each distance metric, we show the results of all AlloyFL techniques per real fault. **Avg**, **Med** and **Std** show the average, median and standard deviation of the corresponding distance metric for each AlloyFL technique over all real faults. **Win** shows the number of times the corresponding AlloyFL technique gives the best distance metric result among all techniques. **Co**, **Un**, **Su**, **Mu** and **Hy** represent AlloyFL$_{co}$, AlloyFL$_{un}$, AlloyFL$_{su}$, AlloyFL$_{mu}$ and AlloyFL$_{hy}$, respectively. We can see that AlloyFL$_{hy}$ has the smallest distance metric result in terms of both **Avg** and **Med**, indicating that AlloyFL$_{hy}$ is the most accurate technique in terms of distance metrics for real faults. Moreover, AlloyFL$_{hy}$ is more stable because it has the smallest **Std**. Out of 38 real faults, AlloyFL$_{hy}$ is the most accurate technique in 25 times under NNUD when $k = 1$, 25 times under NND, and 33 times under NNDW. For NNUD when $k = 5, 10$, AlloyFL$_{hy}$ gives the best results in 26 and

| Model | top1 | | | | | top5 | | | | | top10 | | | | |
|---|---|---|---|---|---|---|---|---|---|---|---|---|---|---|---|
| | Co | Un | Su | Mu | Hy | Co | Un | Su | Mu | Hy | Co | Un | Su | Mu | Hy |
| addr1 | 0 | 0 | 0 | 1 | 1 | 0 | 0 | 0 | 1 | 1 | 0 | 0 | 0 | 1 | 1 |
| arr1 | 0 | 0 | 0 | 0 | 0 | 0 | 0 | 0 | 0 | 0 | 0 | 0 | 0 | 0 | 0 |
| arr2 | 0 | 1 | 1 | 1 | 1 | 0 | 1 | 1 | 1 | 1 | 0 | 1 | 1 | 1 | 1 |
| bst1 | 0 | 0 | 0 | 0 | 0 | 0 | 0 | 0 | 1 | 1 | 0 | 0 | 0 | 1 | 1 |
| bst2 | 0 | 0 | 0 | 0 | 1 | 0 | 0 | 0 | 0 | 1 | 0 | 0 | 0 | 1 | 1 |
| bst3 | 0 | 0 | 0 | 1 | 1 | 0 | 0 | 0 | 2 | 1 | 0 | 0 | 0 | 5 | 1 |
| bempl1 | 0 | 0 | 0 | 0 | 0 | 0 | 0 | 0 | 0 | 0 | 0 | 0 | 0 | 0 | 0 |
| cd1 | 0 | 0 | 0 | 0 | 0 | 0 | 0 | 0 | 2 | 2 | 0 | 0 | 0 | 3 | 2 |
| cd2 | 0 | 0 | 0 | 1 | 1 | 0 | 0 | 0 | 1 | 1 | 0 | 0 | 0 | 1 | 1 |
| ctree1 | 0 | 0 | 0 | 0 | 0 | 1 | 0 | 0 | 0 | 0 | 1 | 0 | 0 | 0 | 0 |
| dll1 | 0 | 0 | 0 | 0 | 0 | 0 | 0 | 0 | 0 | 0 | 0 | 0 | 0 | 0 | 0 |
| dll2 | 0 | 0 | 0 | 1 | 1 | 0 | 0 | 0 | 2 | 1 | 0 | 0 | 0 | 2 | 2 |
| dll3 | 1 | 0 | 0 | 0 | 0 | 1 | 0 | 0 | 2 | 2 | 1 | 0 | 0 | 5 | 4 |
| dll4 | 0 | 0 | 0 | 1 | 1 | 0 | 0 | 0 | 1 | 1 | 0 | 0 | 0 | 1 | 1 |
| farmer1 | 0 | 0 | 0 | 0 | 0 | 0 | 0 | 0 | 0 | 0 | 0 | 0 | 0 | 0 | 0 |
| fsm1 | 0 | 0 | 0 | 1 | 1 | 0 | 0 | 0 | 1 | 1 | 0 | 0 | 0 | 1 | 1 |
| fsm2 | 0 | 0 | 0 | 1 | 1 | 0 | 0 | 0 | 1 | 1 | 0 | 0 | 0 | 1 | 1 |
| grade1 | 1 | 0 | 0 | 0 | 0 | 1 | 0 | 0 | 0 | 1 | 1 | 0 | 0 | 0 | 1 |
| other1 | 0 | 0 | 0 | 0 | 0 | 0 | 0 | 0 | 1 | 1 | 0 | 0 | 0 | 2 | 2 |
| stu1 | 0 | 0 | 0 | 0 | 0 | 0 | 0 | 0 | 0 | 0 | 0 | 0 | 0 | 0 | 0 |
| stu2 | 0 | 0 | 0 | 0 | 0 | 0 | 0 | 0 | 1 | 2 | 0 | 0 | 0 | 2 | 2 |
| stu3 | 0 | 1 | 1 | 0 | 0 | 0 | 2 | 2 | 0 | 0 | 0 | 2 | 2 | 1 | 1 |
| stu4 | 0 | 0 | 0 | 0 | 0 | 0 | 0 | 0 | 0 | 0 | 0 | 0 | 0 | 0 | 0 |
| stu5 | 0 | 0 | 0 | 0 | 0 | 0 | 0 | 0 | 0 | 0 | 0 | 0 | 0 | 0 | 0 |
| stu6 | 0 | 0 | 0 | 1 | 1 | 0 | 0 | 0 | 2 | 1 | 1 | 0 | 0 | 2 | 2 |
| stu7 | 0 | 1 | 1 | 1 | 1 | 0 | 2 | 2 | 2 | 2 | 0 | 2 | 2 | 3 | 2 |
| stu8 | 0 | 0 | 0 | 0 | 0 | 0 | 0 | 0 | 0 | 0 | 0 | 0 | 0 | 0 | 0 |
| stu9 | 0 | 0 | 0 | 1 | 1 | 0 | 0 | 0 | 1 | 1 | 0 | 0 | 0 | 1 | 1 |
| stu10 | 0 | 0 | 0 | 0 | 0 | 0 | 0 | 0 | 0 | 0 | 0 | 0 | 0 | 0 | 0 |
| stu11 | 0 | 0 | 0 | 0 | 0 | 0 | 0 | 0 | 0 | 0 | 0 | 0 | 0 | 0 | 0 |
| stu12 | 0 | 0 | 0 | 0 | 0 | 0 | 0 | 0 | 2 | 2 | 0 | 0 | 0 | 2 | 2 |
| stu13 | 0 | 0 | 0 | 1 | 1 | 0 | 0 | 0 | 1 | 1 | 0 | 0 | 0 | 1 | 1 |
| stu14 | 0 | 0 | 0 | 1 | 1 | 0 | 0 | 0 | 2 | 1 | 1 | 0 | 0 | 2 | 1 |
| stu15 | 0 | 0 | 0 | 0 | 0 | 0 | 0 | 0 | 0 | 0 | 0 | 0 | 0 | 0 | 0 |
| stu16 | 1 | 0 | 0 | 0 | 1 | 1 | 0 | 0 | 0 | 1 | 3 | 0 | 0 | 0 | 1 |
| stu17 | 0 | 0 | 0 | 0 | 0 | 0 | 0 | 0 | 0 | 0 | 1 | 0 | 0 | 1 | 0 |
| stu18 | 0 | 0 | 0 | 0 | 0 | 0 | 1 | 1 | 1 | 1 | 0 | 1 | 1 | 1 | 1 |
| stu19 | 0 | 1 | 1 | 1 | 1 | 0 | 2 | 2 | 2 | 2 | 0 | 2 | 2 | 3 | 3 |
| **Sum** | 3 | 4 | 4 | 14 | 16 | 4 | 8 | 8 | 30 | 30 | 9 | 8 | 8 | 44 | 37 |
| **Win** | 3 | 4 | 4 | 14 | 16 | 3 | 5 | 5 | 20 | 20 | 4 | 3 | 3 | 23 | 18 |

**Figure 8: Top-k Metrics for Real Faults.**

15 times, respective, which is close to AlloyFL$_{mu}$. AlloyFL$_{mu}$ is slightly less accurate than AlloyFL$_{hy}$ in terms of **Avg** and **Med**. AlloyFL$_{co}$, AlloyFL$_{un}$ and AlloyFL$_{su}$ perform almost equally bad in terms of **Avg** under NNUD metrics, and AlloyFL$_{un}$ is even worse than AlloyFL$_{co}$ and AlloyFL$_{su}$ under NND and NNDW metrics. All of AlloyFL$_{co}$, AlloyFL$_{un}$ and AlloyFL$_{su}$ are significantly worse than AlloyFL$_{mu}$ and AlloyFL$_{hy}$. AlloyFL$_{co}$ is accurate for omission faults which happen at the level of paragraph bodies, e.g. when users leave the entire predicate body empty (**stu16**) or miss some conjunct/disjunct constraints at the body of a predicate (**grade1**). On the contrary, AlloyFL$_{mu}$ is not accurate for omission errors because no mutation operator is applicable for an omitted faulty expression/formula. As a consequence, we design AlloyFL$_{hy}$ to leverage the benefits from both AlloyFL$_{co}$ and AlloyFL$_{mu}$. AlloyFL$_{un}$ prioritizes AST nodes that are highlighted the most number of times by the unsat core across all unsatisfiable failing tests and is designed to be comparable or more accurate than using a single unsatisfiable failing test, i.e. the traditional way an Alloy user would debug a faulty model using the unsat core. Our experiments show that the unsat core's accuracy in highlighting suspicious Alloy code is comparable to SBFL (AlloyFL$_{co}$) and significantly worse than MBFL (AlloyFL$_{mu}$ and AlloyFL$_{hy}$).

Figure 8 shows the traditional top-k metric results, i.e. the number of top k suspicious nodes that exactly match the faulty nodes, of AlloyFL for real faults. We highlighted the most accurate AlloyFL techniques w.r.t. each top-k metric in bold. **Sum** shows the total number of faults that exactly match the top-k suspicious nodes for each AlloyFL technique over all real faults. **Win** shows the number of times the corresponding AlloyFL technique gives the best top-k metric result among all techniques. Similar to the observation for distance metrics, AlloyFL$_{mu}$ and AlloyFL$_{hy}$ perform equally well and are significantly more accurate than AlloyFL$_{co}$, AlloyFL$_{un}$ and AlloyFL$_{su}$. AlloyFL$_{hy}$ locates 2 more faulty AST nodes than AlloyFL$_{mu}$ for top-1 metric but it locates 7 less faulty AST nodes than AlloyFL$_{mu}$ for top-10 metric. Both AlloyFL$_{hy}$ and AlloyFL$_{mu}$ locate the same number (but different set) of faulty AST nodes in total for top-5 metric. AlloyFL$_{co}$, AlloyFL$_{un}$ and AlloyFL$_{su}$ are comparable to each other and they locate more or less the same number of faulty AST nodes for top-k metrics (except that AlloyFL$_{un}$ and AlloyFL$_{su}$ locate 4 more faulty AST nodes than AlloyFL$_{co}$ for top-5 metric).

Figure 9 shows the distance metric results of AlloyFL for mutant faults. We use Ochiai formula for AlloyFL$_{co}$, AlloyFL$_{mu}$ and AlloyFL$_{hy}$. The most accurate techniques w.r.t. each distance metric are highlighted in bold. **Avg**, **Med** and **Std** show the average, median and standard deviation of the corresponding distance metric for each AlloyFL technique over all 9000 mutant faults. Each row shows the distance metric results for various AlloyFL techniques on average over 500 second-order mutant faults. AlloyFL$_{mu}$ is the most accurate technique in terms of **Avg**, **Med** and **Std** under the NNUD metrics ($k = 1, 5, 10$). AlloyFL$_{hy}$ is the most accurate technique in terms of **Avg**, **Med** and **Std** under the NND and NNDW metrics. Both AlloyFL$_{mu}$ and AlloyFL$_{hy}$ significantly outperform AlloyFL$_{co}$, AlloyFL$_{un}$ and AlloyFL$_{su}$. AlloyFL$_{co}$ is the least accurate technique in terms of **Avg** under NNUD metrics ($k = 1, 5, 10$), and AlloyFL$_{un}$ is the least accurate technique under NND and NNDW metrics.

Figure 10 shows the traditional top-k metric results of AlloyFL for mutant faults. We highlighted the most accurate AlloyFL techniques w.r.t. each top-k metric in bold. **Sum** shows the sum of the average faults (over 500 mutants) that exactly match the top-k suspicious nodes for each AlloyFL technique over all 18 unique models. **Win** shows the number of times the corresponding AlloyFL technique gives the best top-k metric result among all techniques. Similar to the observation for distance metrics, AlloyFL$_{mu}$ and AlloyFL$_{hy}$ are equally accurate and significantly better than AlloyFL$_{co}$, AlloyFL$_{un}$ and AlloyFL$_{su}$. AlloyFL$_{co}$ is the least accurate technique for top-1 metric and AlloyFL$_{un}$ is the least accurate technique for top-5 and top-10 metric.

Overall, both AlloyFL$_{mu}$ and AlloyFL$_{hy}$ are significantly more accurate than AlloyFL$_{co}$, AlloyFL$_{un}$ and AlloyFL$_{su}$ for both real faults and mutant faults. AlloyFL$_{su}$ is comparable or more accurate than both AlloyFL$_{co}$ and AlloyFL$_{un}$. AlloyFL$_{co}$ and AlloyFL$_{un}$ are the least accurate techniques and they are comparable to each other. AlloyFL$_{un}$ gives better result for NNUD metrics and worse result for NND and NNDW metrics, compared to AlloyFL$_{co}$. All of AlloyFL$_{co}$, AlloyFL$_{un}$ and AlloyFL$_{su}$ are comparable in terms of top-k metrics.

Figure 9: Distance Metrics for Mutant Faults.

| Model | nnud1 | | | | | nnud5 | | | | | nnud10 | | | | | nnd | | | | | nndw | | | | |
|---|---|---|---|---|---|---|---|---|---|---|---|---|---|---|---|---|---|---|---|---|---|---|---|---|---|
| | Co | Un | Su | Mu | Hy | Co | Un | Su | Mu | Hy | Co | Un | Su | Mu | Hy | Co | Un | Su | Mu | Hy | Co | Un | Su | Mu | Hy |
| addr | 32.3 | 32.0 | 32.0 | **5.5** | 10.0 | 30.4 | 31.9 | 30.5 | **7.4** | 13.2 | 27.5 | 28.6 | 28.4 | **10.4** | 12.7 | 21.1 | 21.0 | 19.7 | **5.6** | 12.5 | 22.9 | 38.4 | 31.7 | **8.3** | 13.7 |
| array | 19.4 | 7.0 | 7.5 | **2.4** | 3.2 | 18.1 | 8.1 | 8.8 | **5.7** | 6.1 | 18.8 | 8.1 | 9.6 | **7.7** | 9.6 | 14.7 | 6.9 | 5.9 | **2.7** | 2.9 | 21.0 | 16.4 | 7.8 | 5.1 | **4.2** |
| bst | 59.8 | 43.6 | 42.4 | **4.3** | 7.5 | 64.1 | 50.9 | 47.9 | **6.8** | 7.2 | 71.9 | 50.7 | 51.4 | **10.8** | 10.9 | 23.5 | 25.5 | 19.1 | 4.4 | **3.7** | 31.1 | 51.7 | 27.7 | 6.4 | **4.3** |
| bempl | 13.0 | 19.2 | 17.2 | **5.7** | 7.7 | 8.8 | 16.6 | 12.6 | **6.5** | 7.7 | 11.3 | 16.6 | 14.1 | **7.0** | 9.8 | 4.7 | 16.3 | 10.2 | **3.4** | 3.9 | 5.1 | 46.7 | 20.8 | 4.6 | **4.6** |
| bt | 23.8 | 12.9 | 12.3 | **3.1** | 3.5 | 28.3 | 17.0 | 17.6 | 5.8 | **5.7** | 28.3 | 17.0 | 17.8 | **8.8** | 9.4 | 12.5 | 9.4 | 8.8 | 2.7 | **2.5** | 17.5 | 16.5 | 12.8 | 3.8 | **3.3** |
| cd | 13.0 | 13.6 | 13.5 | 3.1 | **1.8** | 16.0 | 16.4 | 15.3 | 5.6 | **5.0** | 17.1 | 16.4 | 16.2 | **8.1** | 8.4 | 4.9 | 10.3 | 6.1 | 3.0 | **1.5** | 6.0 | 20.9 | 9.6 | 4.4 | **1.7** |
| ctree | 29.1 | 17.9 | 17.8 | **5.9** | 9.2 | 24.2 | 24.6 | 22.8 | **7.3** | 11.1 | 27.4 | 24.6 | 24.7 | **9.6** | 11.9 | 16.1 | 13.5 | 10.7 | **4.7** | 6.8 | 19.0 | 28.3 | 16.4 | **6.3** | 7.9 |
| dijkstra | 46.5 | 49.5 | 53.0 | 6.6 | **5.1** | 58.6 | 46.1 | 44.0 | 9.6 | **7.1** | 71.0 | 48.9 | 46.2 | 13.3 | **11.8** | 18.0 | 37.5 | 23.9 | 12.0 | **6.8** | 59.8 | 127.2 | 62.2 | 26.1 | **12.6** |
| dll | 22.8 | 18.5 | 19.1 | 4.2 | **3.4** | 27.7 | 28.3 | 25.7 | 5.8 | **5.3** | 30.2 | 28.3 | 28.6 | **8.9** | 10.0 | 10.4 | 14.2 | 12.5 | 3.4 | **2.1** | 13.0 | 33.9 | 19.5 | 4.8 | **2.3** |
| farmer | 45.3 | 29.2 | 29.1 | **13.1** | 15.2 | 41.7 | 40.5 | 39.2 | **14.4** | 19.4 | 38.2 | 40.0 | 39.2 | **16.6** | 21.5 | 27.5 | 24.0 | 23.1 | 12.1 | **10.0** | 35.1 | 67.9 | 49.2 | 31.3 | **12.7** |
| fsm | 27.9 | 16.1 | 15.9 | 11.1 | **10.2** | 24.2 | 21.1 | 20.2 | **12.3** | 12.9 | 25.5 | 22.0 | 22.4 | **13.9** | 15.7 | 21.6 | 11.8 | 10.8 | 10.7 | **8.4** | 27.3 | 28.2 | 16.5 | 16.8 | **10.0** |
| fullTree | 27.5 | 17.8 | 17.6 | 4.1 | **3.4** | 34.5 | 23.5 | 23.6 | 6.3 | **5.5** | 34.8 | 23.5 | 24.1 | **9.6** | 9.7 | 11.2 | 13.0 | 10.6 | 3.6 | **2.3** | 15.6 | 26.7 | 15.9 | 4.8 | **2.8** |
| grade | 12.1 | 19.3 | 18.4 | 2.2 | **1.9** | 10.7 | 14.6 | 12.8 | **4.8** | 5.3 | 13.4 | 14.6 | 14.3 | **6.6** | 8.8 | 5.3 | 12.9 | 10.0 | 2.5 | **1.9** | 7.6 | 48.0 | 31.1 | 3.3 | **2.1** |
| hshake | 38.7 | 25.8 | 26.6 | **6.1** | 7.0 | 38.2 | 33.7 | 32.5 | **8.2** | 8.3 | 40.5 | 35.1 | 35.1 | **11.8** | 13.1 | 24.6 | 21.7 | 18.2 | 6.6 | **5.2** | 35.8 | 46.2 | 26.4 | 11.2 | **6.5** |
| nqueens | 22.5 | 5.0 | 5.1 | **3.0** | 3.5 | 26.4 | 7.5 | 7.6 | 6.0 | **5.7** | 26.4 | 7.6 | 8.5 | 9.4 | **10.0** | 15.8 | 5.2 | 3.8 | 3.8 | **3.4** | 31.4 | 15.2 | **6.1** | 6.9 | 8.9 |
| other | 10.3 | 10.6 | 11.3 | **2.5** | 2.8 | 8.0 | 11.3 | 9.9 | **4.9** | 5.8 | 10.3 | 11.3 | 11.6 | **6.7** | 9.9 | 4.4 | 10.7 | 8.2 | 2.4 | **1.8** | 4.8 | 24.0 | 14.3 | 3.9 | **1.9** |
| sll | 9.1 | 11.9 | 11.0 | **2.3** | 2.5 | 11.7 | 13.6 | 13.3 | **4.5** | 5.0 | 11.7 | 13.6 | 13.3 | **6.2** | 7.3 | 4.9 | 11.5 | 9.2 | 2.1 | **2.1** | 7.6 | 22.6 | 17.2 | **2.5** | 2.5 |
| stu | 45.9 | 30.8 | 34.9 | **5.2** | 6.0 | 50.1 | 48.4 | 43.8 | **6.7** | 7.6 | 53.8 | 49.9 | 46.5 | 10.8 | **10.4** | 15.1 | 21.9 | 21.4 | 4.8 | **3.4** | 21.1 | 52.9 | 32.5 | 8.1 | **3.8** |
| **Avg** | 27.7 | 21.2 | 21.4 | 5.0 | 5.8 | 29.0 | 25.2 | 23.8 | 7.1 | 8.0 | 31.0 | 25.4 | 25.1 | 9.8 | 11.2 | 14.2 | 16.0 | 12.9 | 5.0 | 4.5 | 21.2 | 39.5 | 23.2 | 8.8 | 5.9 |
| **Med** | 25.6 | 18.2 | 17.7 | 4.2 | 4.3 | 27.1 | 22.3 | 21.5 | 6.4 | 6.6 | 27.4 | 22.8 | 23.2 | 9.5 | 10.0 | 14.9 | 13.3 | 10.6 | 3.7 | 3.4 | 20.0 | 31.1 | 18.4 | 5.7 | 4.2 |
| **Std** | 14.2 | 11.7 | 12.3 | 2.9 | 3.5 | 16.1 | 13.5 | 12.7 | 2.5 | 3.7 | 18.1 | 13.8 | 13.2 | 2.7 | 3.1 | 7.3 | 7.7 | 6.1 | 3.2 | 3.1 | 13.7 | 25.8 | 14.0 | 7.8 | 4.0 |
| **Win** | 0 | 0 | 0 | 12 | 6 | 0 | 0 | 0 | 12 | 6 | 0 | 1 | 0 | 15 | 2 | 0 | 0 | 0 | 4 | 14 | 0 | 0 | 1 | 3 | 14 |

| Model | top1 | | | | | top5 | | | | | top10 | | | | |
|---|---|---|---|---|---|---|---|---|---|---|---|---|---|---|---|
| | Co | Un | Su | Mu | Hy | Co | Un | Su | Mu | Hy | Co | Un | Su | Mu | Hy |
| addr | 0.1 | 0.1 | 0.1 | **0.8** | 0.6 | 0.5 | 0.2 | 0.3 | **1.4** | 0.9 | 0.5 | 0.2 | 0.3 | **1.5** | 1.4 |
| array | 0.1 | 0.6 | 0.6 | 0.8 | **0.8** | 0.4 | 0.8 | 0.9 | 1.2 | **1.3** | 0.4 | 0.8 | 0.9 | 1.5 | **1.6** |
| bst | 0.1 | 0.2 | 0.2 | **0.7** | 0.7 | 0.1 | 0.2 | 0.3 | **1.1** | 1.0 | 0.1 | 0.2 | 0.3 | **1.2** | 1.2 |
| bempl | 0.2 | 0.1 | 0.1 | **0.8** | 0.6 | 0.9 | 0.1 | 0.6 | **1.1** | 1.0 | 1.1 | 0.1 | 0.8 | 1.4 | **1.6** |
| bt | 0.1 | 0.4 | 0.4 | **0.7** | 0.7 | 0.1 | 0.4 | 0.4 | 1.1 | **1.1** | 0.1 | 0.4 | 0.4 | 1.3 | **1.3** |
| cd | 0.2 | 0.2 | 0.3 | 0.8 | **0.8** | 0.4 | 0.3 | 0.4 | 1.3 | **1.3** | 0.4 | 0.3 | 0.4 | 1.4 | **1.5** |
| ctree | 0.1 | 0.2 | 0.2 | **0.7** | 0.6 | 0.4 | 0.2 | 0.3 | **1.0** | 0.9 | 0.4 | 0.2 | 0.3 | 1.1 | **1.2** |
| dijkstra | 0.1 | 0.3 | 0.3 | 0.8 | **0.8** | 0.2 | 0.4 | 0.4 | **1.2** | 1.1 | 0.2 | 0.4 | 0.5 | 1.3 | **1.3** |
| dll | 0.1 | 0.1 | 0.2 | 0.7 | **0.7** | 0.1 | 0.1 | 0.2 | **1.3** | 1.2 | 0.1 | 0.1 | 0.2 | 1.4 | **1.5** |
| farmer | 0.1 | 0.1 | 0.1 | **0.6** | 0.5 | 0.3 | 0.2 | 0.2 | **0.8** | 0.7 | 0.4 | 0.2 | 0.2 | 0.9 | **1.0** |
| fsm | 0.1 | 0.3 | 0.3 | 0.4 | **0.4** | 0.3 | 0.3 | 0.4 | 0.8 | **0.8** | 0.3 | 0.3 | 0.4 | 0.9 | **1.0** |
| fullTree | 0.1 | 0.3 | 0.3 | **0.8** | 0.8 | 0.2 | 0.3 | 0.3 | 1.1 | **1.1** | 0.2 | 0.3 | 0.3 | 1.3 | **1.3** |
| grade | 0.2 | 0.1 | 0.2 | **0.9** | 0.9 | 0.7 | 0.2 | 0.3 | **1.4** | 1.4 | 0.8 | 0.2 | 0.4 | 1.5 | **1.7** |
| hshake | 0.1 | 0.2 | 0.2 | 0.7 | **0.7** | 0.4 | 0.2 | 0.3 | **1.0** | 1.0 | 0.3 | 0.2 | 0.3 | 1.0 | **1.1** |
| nqueens | 0.1 | **0.8** | 0.8 | 0.8 | 0.7 | 0.1 | 1.1 | 1.1 | **1.3** | 1.2 | 0.1 | 1.1 | 1.1 | 1.4 | **1.5** |
| other | 0.2 | 0.2 | 0.2 | **0.9** | 0.9 | 0.9 | 0.3 | 0.3 | 1.2 | **1.4** | 1.1 | 0.3 | 0.8 | 1.3 | **1.7** |
| sll | 0.3 | 0.0 | 0.1 | **0.8** | 0.7 | 0.5 | 0.0 | 0.1 | 1.4 | **1.4** | 0.5 | 0.0 | 0.1 | 1.5 | **1.5** |
| stu | 0.1 | 0.1 | 0.1 | **0.8** | 0.8 | 0.2 | 0.2 | 0.2 | **1.1** | 1.1 | 0.3 | 0.2 | 0.2 | **1.3** | 1.3 |
| **Sum** | 2.1 | 4.3 | 4.5 | 13.2 | 12.8 | 6.7 | 5.6 | 7.4 | 20.7 | 20.2 | 7.5 | 5.6 | 8.1 | 23.2 | 24.6 |
| **Win** | 0 | 1 | 1 | 10 | 7 | 0 | 0 | 0 | 11 | 8 | 0 | 0 | 0 | 3 | 15 |

Figure 10: Top-k Metrics for Mutant Faults.

MBFL techniques (e.g. AlloyFL$_{mu}$ and AlloyFL$_{hy}$) are the most accurate fault localization techniques and are significantly better than SBFL techniques (e.g. AlloyFL$_{co}$) and SAT-based techniques (e.g. AlloyFL$_{un}$ and AlloyFL$_{su}$). Importantly, our result indicates that AlloyFL$_{mu}$ and AlloyFL$_{hy}$ can more accurately highlight suspicious Alloy code compared to state-of-the-art unsat core.

Because of space limit, we cannot show the time overhead of AlloyFL for individual faults. On average, AlloyFL$_{co}$, AlloyFL$_{un}$ and AlloyFL$_{su}$ take less than 5 sec to run for both mutant faults and real faults. AlloyFL$_{mu}$ takes on average 24.2 sec for mutant faults and 33.7 sec for real faults. AlloyFL$_{hy}$ takes on average 38.0 sec for mutant faults and 67.0 sec for real faults. Both AlloyFL$_{mu}$ and AlloyFL$_{hy}$ run the test for each mutation and AlloyFL$_{hy}$ has the extra overhead to run AlloyFL$_{co}$ and compute the average suspiciousness scores, thus AlloyFL$_{hy}$ is slower than AlloyFL$_{mu}$ and both of them are slower than AlloyFL$_{co}$, AlloyFL$_{un}$ and AlloyFL$_{su}$.

| | Formula | nnud1 | nnud5 | nnud10 | nnd | nndw | top1 | top5 | top10 |
|---|---|---|---|---|---|---|---|---|---|
| Co | Tarantula | 25.6 | 32.8 | **38.8** | 15.0 | **20.9** | **0.1** | **0.2** | **0.3** |
| | Ochiai | **22.4** | **31.7** | 38.9 | **14.9** | 21.7 | **0.1** | 0.1 | 0.2 |
| | Op2 | 28.8 | 35.7 | 38.9 | 23.6 | 32.1 | 0.0 | 0.1 | 0.2 |
| | Barinel | 25.6 | 32.8 | **38.8** | 15.0 | **20.9** | **0.1** | **0.2** | **0.3** |
| | DStar | 23.4 | 33.0 | 38.9 | 16.5 | 23.5 | **0.1** | 0.1 | 0.2 |
| Mu | Tarantula | 10.2 | 15.2 | 19.0 | 8.7 | 13.3 | 0.3 | 0.8 | 1.0 |
| | Ochiai | 8.2 | **11.4** | **15.9** | 7.5 | **11.5** | **0.4** | **0.8** | **1.2** |
| | Op2 | 12.6 | 12.8 | 17.4 | 10.4 | 15.4 | 0.3 | 0.7 | 1.0 |
| | Barinel | 10.2 | 15.2 | 19.0 | 8.7 | 13.3 | 0.3 | 0.8 | 1.0 |
| | DStar | **8.2** | **11.4** | 16.3 | **7.8** | 12.2 | 0.3 | 0.7 | 1.1 |
| Hy | Tarantula | 9.9 | 12.4 | **14.5** | 5.9 | 8.1 | 0.4 | **0.8** | 1.1 |
| | Ochiai | **6.1** | **9.1** | 14.6 | **4.7** | **7.1** | **0.4** | 0.8 | 1.0 |
| | Op2 | 26.2 | 22.5 | 19.2 | 21.8 | 26.7 | 0.3 | 0.3 | 0.7 |
| | Barinel | 10.0 | 12.4 | **14.5** | 6.1 | 8.2 | 0.3 | **0.8** | **1.1** |
| | DStar | 6.6 | 9.2 | 15.0 | 5.1 | 7.8 | 0.3 | 0.7 | 0.9 |

Figure 11: Formula Impact on AlloyFL for Real Faults.

MBFL techniques are significantly slower than SBFL techniques and SAT-based techniques. But since all techniques finish under 2 min on average, the time overhead are reasonable.

## 5.3 RQ2: Suspiciousness Formula Impact

Since AlloyFL$_{un}$ and AlloyFL$_{su}$ do not use suspiciousness formulas, we answer *RQ2* only for AlloyFL$_{co}$, AlloyFL$_{mu}$ and AlloyFL$_{hy}$.

Figure 11 shows the average results for both distance metrics and top-k metrics under different suspiciousness formulas for AlloyFL$_{co}$, AlloyFL$_{mu}$ and AlloyFL$_{hy}$ over 38 real faults. The best results for each metric among all suspiciousness formulas are highlighted in bold. We can see that for AlloyFL$_{co}$, the metric values do not change much for various formulas and Op2 seems to be the worst formula. For AlloyFL$_{mu}$, Ochiai and DStar outperform other formulas and Ochiai is slightly better than DStar. For AlloyFL$_{hy}$, Ochiai seems to be comparable or better than other formulas, followed by DStar. Although Tarantula and Barinel are sometimes the best formulas to use, the improvement is not significant over Ochiai.

| | Formula | nnud1 | nnud5 | nnud10 | nnd | nndw | top1 | top5 | top10 |
|---|---|---|---|---|---|---|---|---|---|
| Co | Tarantula | 29.2 | **29.0** | 31.0 | 16.3 | 23.1 | 0.1 | 0.4 | 0.4 |
| | Ochiai | **27.7** | 29.0 | 31.0 | **14.2** | **21.2** | 0.1 | 0.4 | 0.4 |
| | Op2 | 29.9 | 29.5 | 31.2 | 15.8 | 23.0 | **0.1** | 0.4 | 0.4 |
| | Barinel | 29.2 | **29.0** | 31.0 | 16.3 | 23.1 | 0.1 | 0.4 | 0.4 |
| | DStar | 27.8 | 29.1 | 31.0 | 14.4 | 21.4 | 0.1 | 0.4 | **0.4** |
| Mu | Tarantula | 10.8 | 9.8 | 11.9 | 8.0 | 12.3 | 0.5 | 1.0 | 1.1 |
| | Ochiai | **5.0** | **7.1** | **9.8** | **5.0** | **8.8** | **0.7** | **1.2** | **1.3** |
| | Op2 | 7.3 | 8.5 | 10.6 | 6.6 | 11.0 | 0.7 | 1.1 | 1.2 |
| | Barinel | 10.8 | 9.8 | 11.9 | 8.0 | 12.3 | 0.5 | 1.0 | 1.1 |
| | DStar | 5.4 | 7.5 | 9.9 | 5.3 | 9.2 | 0.7 | 1.1 | 1.3 |
| Hy | Tarantula | 10.8 | 9.8 | 12.8 | 7.1 | 8.4 | 0.5 | 1.1 | 1.3 |
| | Ochiai | **5.8** | **8.0** | **11.2** | **4.5** | **5.9** | **0.7** | 1.1 | **1.4** |
| | Op2 | 17.0 | 13.5 | 12.7 | 11.8 | 14.6 | 0.4 | 0.9 | 1.3 |
| | Barinel | 11.0 | 9.7 | 12.8 | 7.0 | 8.2 | 0.5 | 1.1 | 1.3 |
| | DStar | 6.4 | 8.5 | 11.3 | 5.2 | 6.7 | 0.7 | 1.1 | 1.3 |

**Figure 12: Formulas Impact on AlloyFL for Mutant Faults.**

Figure 12 shows the average results for both distance metrics and top-k metrics under different suspiciousness formulas for AlloyFL$_{co}$, AlloyFL$_{mu}$ and AlloyFL$_{hy}$ over 9000 mutant faults. The best results for each metric among all suspiciousness formulas are highlighted in bold. For AlloyFL$_{co}$, all formulas give similar results. For AlloyFL$_{mu}$ and AlloyFL$_{hy}$, Ochiai gives the best results among all metrics and DStar gives the second best results.

Overall, the choice of formulas does not impact the accuracy of AlloyFL$_{co}$ much for both real faults and mutant faults. Ochiai seems to be the best formula to choose for AlloyFL$_{mu}$ and AlloyFL$_{hy}$ (followed by DStar) for both real faults and mutant faults.

> Suspiciousness formulas do not have much impact on the accuracy of SBFL techniques (e.g. AlloyFL$_{co}$). Ochiai formula gives the best result of most metrics for MBFL techniques (e.g. AlloyFL$_{mu}$ and AlloyFL$_{hy}$).

## 5.4 Threats to Validity

There exists several threats to the validity of our results. The real faulty models we use in the experiment are limited in the sense that most of them are written by graduate students. So the experiment results may not generalize to faulty models written by experienced developers. However, we collect our subject faulty models to the best of our ability. The AUnit tests (e.g., the test in Figure 1b) can require some manual effort to create. In this paper, all tests are *automatically generated* using MuAlloy [61] and the expected behavior (`expect 0 or 1`) of each test is automatically verified using the correct model. In practice, users need to manually specify the expected behavior but no manually effort is needed to write test predicates. We choose to use automatically generated test predicates to evaluate AlloyFL because we did not find real faulty models with enough tests. So our result may not generalize to manually written tests. Additionally, although our distance metrics simulate different ways users may inspect code highlighted by AlloyFL, our result may not generalize to new metrics.

## 6 RELATED WORK

AlloyFL presents the first set of *automated* fault localization techniques that leverage multiple test formulas and we show that it is able to highlight suspicious Alloy code more accurate compared to existing unsat cores, with the help of *automatically* generated AUnit tests.

Automated debugging of Alloy models can be traced back to Alloy's early days when highlighting unsat cores in unsatisfiable Alloy formulas was introduced [53]. Moreover, for satisfiable formulas, Alloy's symmetry breaking indirectly supports debugging by allowing the user to inspect fewer instances [12, 22, 43, 52]. More recent work on Amalgam allows the user to ask questions of the form "why a tuple is or is not in a relation" for a chosen instance [42]. While Amalgam provides a useful tool to aid debugging by allowing the user to enhance their understanding of the model by asking a series of questions, the restricted form of the questions limits its effectiveness, e.g., the user cannot ask why certain formulas hold or not, or why certain relations are empty. AlloyFL provided the fault localization engine for ARepair [60], one of the most recent automated debugging techniques for Alloy. Specifically, ARepair uses the mutation-based fault localization component AlloyFL$_{mu}$.

A number of approaches assist users to write correct Alloy models. Montaghami and Rayside [37, 38] enable Alloy users to more easily provide partial instances, which are tangible, expressive example solutions that aid in writing correct, complete models. Sullivan et al. [57] follow the spirit of JUnit and introduce a test automation framework for Alloy by defining test case, test execution and model coverage. AUnit has enabled further test automation efforts for Alloy, ranging from automated test generation to mutation testing [56, 61].

While our focus in this paper is on declarative models written in Alloy, fault localization for imperative languages is a well-studied area. AlloyFL implements spectrum-based, mutation-based, and SAT-based techniques. Among these, spectrum-based techniques [2, 4, 9, 11, 19, 20, 28, 47, 49], are the most widely studied; they focus on collecting execution information, such as statements and methods. Mutation-based fault localization techniques [39, 44] were introduced more recently. They perform mutations on the faulty program to study their impact on the test execution results and determine likely faulty locations. SAT-based techniques use either the minimal satisfiability [13] or the negation of maximal satisfiability [21] to identify suspicious code.

A number of other techniques have also been proposed for fault localization. Comparing program states between passing and failing tests has shown to be highly effective and was pioneered by delta debugging [67, 68], which has led to various other approaches [8, 14, 70]. Statistic-based approaches [31, 64] focus on determining the likelihood of different portions of a program being faulty. Feedback-based debugging [27, 29] is an interactive fault localization approach that utilizes execution traces and user feedback. Program slicing [5, 6, 32] isolates relevant program elements that can trigger the execution traces that lead to errors.

## 7 CONCLUSIONS

This paper introduces AlloyFL, the a set of fault localization techniques for declarative Alloy models. AlloyFL is the first set of techniques that utilize a suite of "*test*" formulas (either automatically generated or manually written) that capture the expected properties of Alloy models and locates faults at the AST node granularity. Moreover, we propose new distance metrics to evaluate AlloyFL (e.g. NNUD, NND and NNDW) and evaluate AlloyFL using a suite of Alloy models including 38 real faults and 9000 mutant faults. We show that our mutation-based techniques (e.g. AlloyFL$_{mu}$ and AlloyFL$_{hy}$) together with Ochiai suspiciousness formula significantly outperform other baseline techniques, including the spectrum-based technique (e.g. AlloyFL$_{co}$) and the SAT-based techniques (e.g. AlloyFL$_{un}$ and AlloyFL$_{su}$).